\pdfoutput=1
\documentclass[12pt,a4paper]{article}
\usepackage{ifthen} 
\newboolean{pdflatex}
\setboolean{pdflatex}{true} 

\newboolean{articletitles}
\setboolean{articletitles}{true} 

\newboolean{uprightparticles}
\setboolean{uprightparticles}{false} 

\newboolean{inbibliography}
\setboolean{inbibliography}{false} 

\usepackage{microtype}
\usepackage{lineno}  
\usepackage{xspace} 
\usepackage{caption} 

\usepackage{graphicx}  
\usepackage{color}
\usepackage{colortbl}
\graphicspath{{./figs/}} 

\usepackage{amsmath} 
\usepackage{amssymb}
\usepackage{amsfonts}
\usepackage{upgreek} 

\newcommand*\patchAmsMathEnvironmentForLineno[1]{%
\expandafter\let\csname old#1\expandafter\endcsname\csname #1\endcsname
\expandafter\let\csname oldend#1\expandafter\endcsname\csname
end#1\endcsname
 \renewenvironment{#1}%
   {\linenomath\csname old#1\endcsname}%
   {\csname oldend#1\endcsname\endlinenomath}%
}
\newcommand*\patchBothAmsMathEnvironmentsForLineno[1]{%
  \patchAmsMathEnvironmentForLineno{#1}%
  \patchAmsMathEnvironmentForLineno{#1*}%
}
\AtBeginDocument{%
\patchBothAmsMathEnvironmentsForLineno{equation}%
\patchBothAmsMathEnvironmentsForLineno{align}%
\patchBothAmsMathEnvironmentsForLineno{flalign}%
\patchBothAmsMathEnvironmentsForLineno{alignat}%
\patchBothAmsMathEnvironmentsForLineno{gather}%
\patchBothAmsMathEnvironmentsForLineno{multline}%
\patchBothAmsMathEnvironmentsForLineno{eqnarray}%
}

\usepackage{hyperref}    
\usepackage[all]{hypcap} 

\def\lhcb {\mbox{LHCb}\xspace}

\def\cdf    {\mbox{CDF}\xspace}
\def\dzero  {\mbox{D0}\xspace}

\def\MagUp {\mbox{\em Mag\kern -0.05em Up}\xspace}

\ifthenelse{\boolean{uprightparticles}}%
{

 \def\Pmu         {\ensuremath{\upmu}\xspace}

 \def\Ppi         {\ensuremath{\uppi}\xspace}

 \def\Ppsi        {\ensuremath{\uppsi}\xspace}

 \def\PDelta      {\ensuremath{\Delta}\xspace}                 
 \def\PXi      {\ensuremath{\Xi}\xspace}                 
 \def\PLambda      {\ensuremath{\Lambda}\xspace}                 
 \def\PSigma      {\ensuremath{\Sigma}\xspace}                 
 \def\POmega      {\ensuremath{\Omega}\xspace}                 
 \def\PUpsilon      {\ensuremath{\Upsilon}\xspace}                 
 
 \def\PB      {\ensuremath{\mathrm{B}}\xspace}                 
                  
 \def\PD      {\ensuremath{\mathrm{D}}\xspace}

 \def\PJ      {\ensuremath{\mathrm{J}}\xspace}                 
 \def\PK      {\ensuremath{\mathrm{K}}\xspace}

 \def\Pb      {\ensuremath{\mathrm{b}}\xspace}                 
 \def\Pc      {\ensuremath{\mathrm{c}}\xspace}

 \def\Pi      {\ensuremath{\mathrm{i}}\xspace}

 \def\Ps      {\ensuremath{\mathrm{s}}\xspace}

}
{

 \def\Pmu         {\ensuremath{\mu}\xspace}

 \def\Ppi         {\ensuremath{\pi}\xspace}

 \def\Ppsi        {\ensuremath{\psi}\xspace}                 
                  
 \mathchardef\PDelta="7101
 \mathchardef\PXi="7104
 \mathchardef\PLambda="7103
 \mathchardef\PSigma="7106
 \mathchardef\POmega="710A
 \mathchardef\PUpsilon="7107
                  
 \def\PB      {\ensuremath{B}\xspace}                 
                  
 \def\PD      {\ensuremath{D}\xspace}

 \def\PJ      {\ensuremath{J}\xspace}                 
 \def\PK      {\ensuremath{K}\xspace}

 \def\Pb      {\ensuremath{b}\xspace}                 
 \def\Pc      {\ensuremath{c}\xspace}

 \def\Pi      {\ensuremath{i}\xspace}

 \def\Ps      {\ensuremath{s}\xspace}

}

\makeatletter
\ifcase \@ptsize \relax
  \newcommand{\miniscule}{\@setfontsize\miniscule{4}{5}}
\or
  \newcommand{\miniscule}{\@setfontsize\miniscule{5}{6}}
\or
  \newcommand{\miniscule}{\@setfontsize\miniscule{5}{6}}
\fi
\makeatother

\DeclareRobustCommand{\optbar}[1]{\shortstack{{\miniscule (\rule[.5ex]{1.25em}{.18mm})}
  \\ [-.7ex] $#1$}}

\def\mup        {{\ensuremath{\Pmu^+}}\xspace}
\def\mumu       {{\ensuremath{\Pmu^+\Pmu^-}}\xspace}

\def\squark    {{\ensuremath{\Ps}}\xspace}

\def\cquark    {{\ensuremath{\Pc}}\xspace}

\def\bquark    {{\ensuremath{\Pb}}\xspace}

\def\pion   {{\ensuremath{\Ppi}}\xspace}

\def\pip    {{\ensuremath{\pion^+}}\xspace}

\def\kaon    {{\ensuremath{\PK}}\xspace}
  \def\Kbar    {{\kern 0.2em\overline{\kern -0.2em \PK}{}}\xspace}

\def\KorKbar    {\kern 0.18em\optbar{\kern -0.18em K}{}\xspace}

\def\Kp      {{\ensuremath{\kaon^+}}\xspace}
\def\Km      {{\ensuremath{\kaon^-}}\xspace}

  \def\Dbar    {{\kern 0.2em\overline{\kern -0.2em \PD}{}}\xspace}
\def\D       {{\ensuremath{\PD}}\xspace}

\def\DorDbar    {\kern 0.18em\optbar{\kern -0.18em D}{}\xspace}
\def\Dz      {{\ensuremath{\D^0}}\xspace}

\def\Dsp     {{\ensuremath{\D^+_\squark}}\xspace}

\def\B       {{\ensuremath{\PB}}\xspace}
\def\Bbar    {{\ensuremath{\kern 0.18em\overline{\kern -0.18em \PB}{}}}\xspace}

\def\BorBbar    {\kern 0.18em\optbar{\kern -0.18em B}{}\xspace}
\def\Bz      {{\ensuremath{\B^0}}\xspace}

\def\Bu      {{\ensuremath{\B^+}}\xspace}

\def\Bp      {{\ensuremath{\Bu}}\xspace}

\def\Bd      {{\ensuremath{\B^0}}\xspace}

\def\Bcp     {{\ensuremath{\B_\cquark^+}}\xspace}

\def\jpsi     {{\ensuremath{{\PJ\mskip -3mu/\mskip -2mu\Ppsi\mskip 2mu}}}\xspace}

  \def\Y#1S{\ensuremath{\PUpsilon{(#1S)}}\xspace}

\def\Lbar        {{\ensuremath{\kern 0.1em\overline{\kern -0.1em\PLambda}}}\xspace}
\def\LorLbar    {\kern 0.18em\optbar{\kern -0.18em \PLambda}{}\xspace}

\def\BF         {{\ensuremath{\cal B}}\xspace}

\newcommand{\decay}[2]{\ensuremath{#1\!\to #2}\xspace}         

\def\to                 {\ensuremath{\rightarrow}\xspace}

\def\AT#1     {\ensuremath{A_{\mathrm{T}}^{#1}}\xspace}           

\def\C#1      {\ensuremath{\mathcal{C}_{#1}}\xspace}                       
\def\Cp#1     {\ensuremath{\mathcal{C}_{#1}^{'}}\xspace}                    
\def\Ceff#1   {\ensuremath{\mathcal{C}_{#1}^{\mathrm{(eff)}}}\xspace}        
\def\Cpeff#1  {\ensuremath{\mathcal{C}_{#1}^{'\mathrm{(eff)}}}\xspace}       
\def\Ope#1    {\ensuremath{\mathcal{O}_{#1}}\xspace}                       
\def\Opep#1   {\ensuremath{\mathcal{O}_{#1}^{'}}\xspace}                    

\newcommand{\tev}{\ifthenelse{\boolean{inbibliography}}{\ensuremath{~T\kern -0.05em eV}\xspace}{\ensuremath{\mathrm{\,Te\kern -0.1em V}}}\xspace}
\newcommand{\gev}{\ensuremath{\mathrm{\,Ge\kern -0.1em V}}\xspace}
\newcommand{\mev}{\ensuremath{\mathrm{\,Me\kern -0.1em V}}\xspace}
\newcommand{\kev}{\ensuremath{\mathrm{\,ke\kern -0.1em V}}\xspace}
\newcommand{\ev}{\ensuremath{\mathrm{\,e\kern -0.1em V}}\xspace}
\newcommand{\gevc}{\ensuremath{{\mathrm{\,Ge\kern -0.1em V\!/}c}}\xspace}
\newcommand{\mevc}{\ensuremath{{\mathrm{\,Me\kern -0.1em V\!/}c}}\xspace}
\newcommand{\gevcc}{\ensuremath{{\mathrm{\,Ge\kern -0.1em V\!/}c^2}}\xspace}
\newcommand{\gevgevcccc}{\ensuremath{{\mathrm{\,Ge\kern -0.1em V^2\!/}c^4}}\xspace}
\newcommand{\mevcc}{\ensuremath{{\mathrm{\,Me\kern -0.1em V\!/}c^2}}\xspace}

\def\mum  {\ensuremath{{\,\upmu\rm m}}\xspace}

\def\invfb   {\ensuremath{\mbox{\,fb}^{-1}}\xspace}

\def\fs   {\ensuremath{\rm \,fs}\xspace}

\def\gsim{{~\raise.15em\hbox{$>$}\kern-.85em
          \lower.35em\hbox{$\sim$}~}\xspace}
\def\lsim{{~\raise.15em\hbox{$<$}\kern-.85em
          \lower.35em\hbox{$\sim$}~}\xspace}

\def\sPlot{\mbox{\em sPlot}\xspace}

\def\pt         {\mbox{$p_{\rm T}$}\xspace}

\def\bcvegpy    {\mbox{\textsc{Bcvegpy}}\xspace}

\def\evtgen     {\mbox{\textsc{EvtGen}}\xspace}

\def\geant      {\mbox{\textsc{Geant4}}\xspace}

\def\photos     {\mbox{\textsc{Photos}}\xspace}

\def\tell1  {TELL1\xspace}
\def\ukl1   {UKL1\xspace}

\newcommand{\eg}{\mbox{\itshape e.g.}\xspace}

\usepackage{cite} 
\usepackage{mciteplus}

\usepackage{longtable} 

\newcommand{\DecayPsiPPH}{\ensuremath{\decay{\Bcp}{\jpsi p \overline{p}\pip}}\xspace}
\newcommand{\DecayPsiH}{\ensuremath{\decay{\Bcp}{\jpsi\pip}}\xspace}
\usepackage{multirow}
\usepackage{booktabs}
\usepackage{lscape}
\begin{document}
\renewcommand{\thefootnote}{\fnsymbol{footnote}}
\setcounter{footnote}{1}
\begin{titlepage}
\pagenumbering{roman}

\vspace*{-1.5cm}
\centerline{\large EUROPEAN ORGANIZATION FOR NUCLEAR RESEARCH (CERN)}
\vspace*{1.5cm}
\hspace*{-0.5cm}
\begin{tabular*}{\linewidth}{lc@{\extracolsep{\fill}}r}
\ifthenelse{\boolean{pdflatex}}
{\vspace*{-2.7cm}\mbox{\!\!\!\includegraphics[width=.14\textwidth]{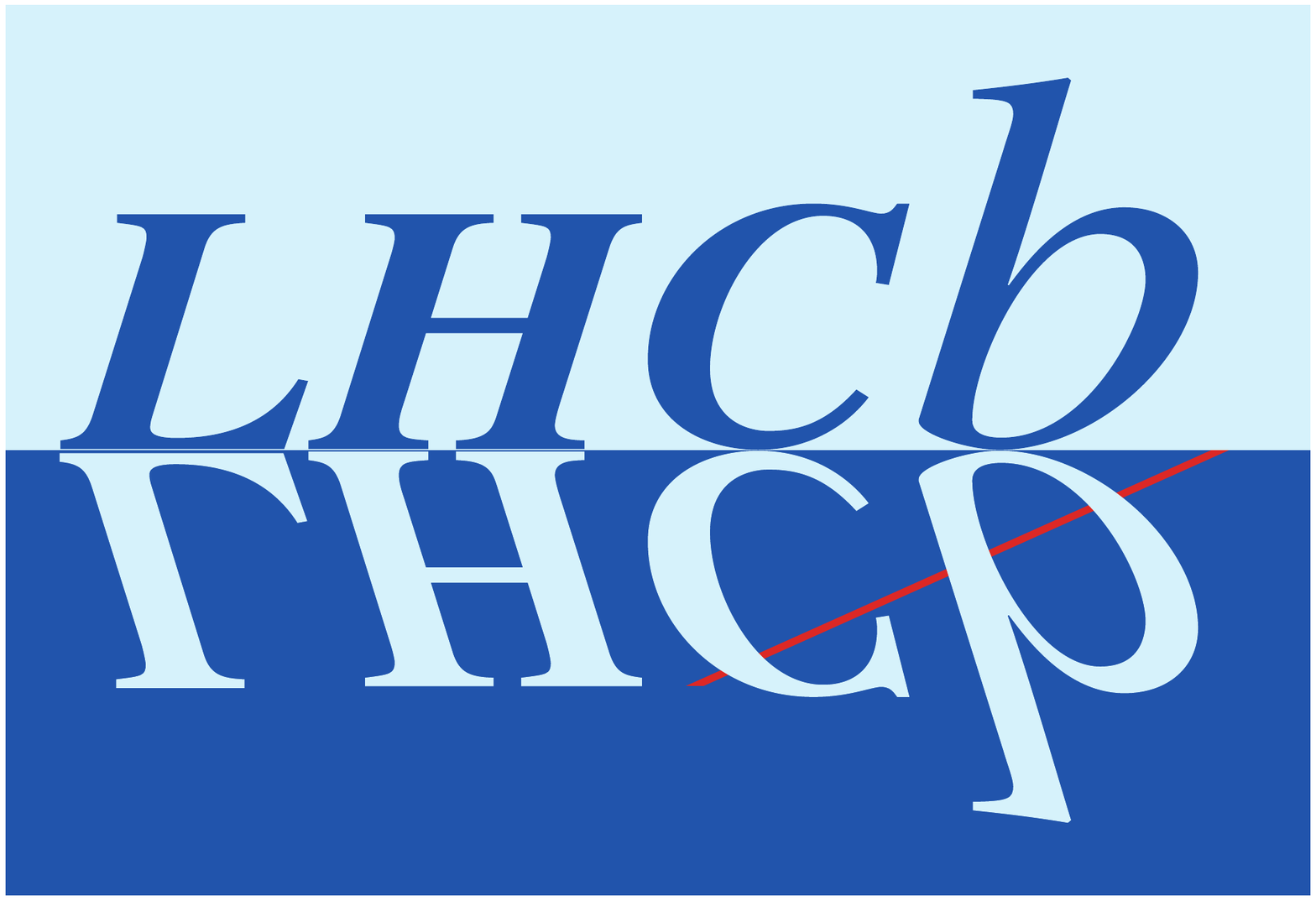}} & &}%
{\vspace*{-1.2cm}\mbox{\!\!\!\includegraphics[width=.12\textwidth]{lhcb-logo.eps}} & &}%
\\
 & & CERN-PH-EP-2014-187\\  
 & & LHCb-PAPER-2014-039\\  
 & & 5 August 2014\\ 
\end{tabular*}

\vspace*{3.0cm}

{\bf\boldmath\huge
\begin{center}
   First observation of a baryonic $B_c^+$ decay
\end{center}
}

\vspace*{1.0cm}

\begin{center}
The LHCb collaboration\footnote{ Authors are listed at the end of this Letter.}
\end{center}

\vspace{\fill}

\begin{abstract}
  \noindent
A baryonic decay of the $\Bcp$ meson, $\DecayPsiPPH$, is observed for the first time, with a 
significance of $7.3$ standard deviations, in $pp$ collision data collected with the LHCb detector and  
corresponding to an integrated luminosity of $3.0\invfb$ taken at center-of-mass energies of $7$ and $8\tev$. 
With the $\DecayPsiH$ decay as normalization channel, the ratio of branching fractions is measured to be
\begin{equation*}
\frac{\BF(\DecayPsiPPH)}{\BF(\DecayPsiH)} = 0.143^{\,+\,0.039}_{\,-\,0.034}\,(\mathrm{stat})\pm0.013\,(\mathrm{syst}).
\end{equation*}
The mass of the $\Bcp$ meson is determined as $M(\Bcp)=6274.0\pm1.8\,(\mathrm{stat})\pm0.4\,(\mathrm{syst})\mevcc$, using the $\DecayPsiPPH$ channel.
\end{abstract}

\vspace*{2.0cm}

\begin{center}
  Submitted to Phys. Rev. Lett.
\end{center}

\vspace{\fill}

{\footnotesize
\centerline{\copyright~CERN on behalf of the \lhcb collaboration, license \href{http://creativecommons.org/licenses/by/4.0/}{CC-BY-4.0}.}}
\vspace*{2mm}

\end{titlepage}


\newpage
\setcounter{page}{2}
\mbox{~}

\cleardoublepage

\renewcommand{\thefootnote}{\arabic{footnote}}
\setcounter{footnote}{0}

\pagestyle{plain} 
\setcounter{page}{1}
\pagenumbering{arabic}

\noindent
The $\Bcp$ meson is the ground state of the $\overline{b}c$ system and is the only doubly heavy flavored meson that
decays weakly (the inclusion of charge conjugated processes is implied throughout this Letter). 
A large number of $\Bcp$ decay modes are expected,
since either the $\overline{b}$ quark or the $c$ quark can decay, with the other quark acting as spectator, 
or the two quarks can annihilate into a virtual $W^+$ boson. 
The $\Bcp$ meson was first observed by \cdf through the semileptonic decay 
\mbox{$\Bcp\to\jpsi l^{+}\nu_lX$}~\cite{Abe:1998wi}, 
and the hadronic decay $\DecayPsiH$ was observed later by \cdf and $\dzero$~\cite{Aaltonen:2007gv,abazov:2008kv}. 
Many more hadronic decay channels of the $\Bcp$ meson have been observed by \lhcb~\cite{LHCb-PAPER-2011-044,LHCb-PAPER-2013-021,LHCb-PAPER-2012-054,LHCb-PAPER-2013-010,LHCb-PAPER-2013-047,LHCb-PAPER-2013-044,Bc2Jpsi5H}. 
At \lhcb, the \Bcp mass was measured in the $\Bcp\to\jpsi\pip$ \cite{LHCb-PAPER-2012-028} and $\Bcp\to\jpsi\Dsp$
\cite{LHCb-PAPER-2013-010} decays,  and its lifetime has been determined using the $\Bcp\to\jpsi\mup\nu_\mu X$
decay~\cite{LHCb-PAPER-2013-063}.
However, baryonic decays of $\Bcp$ mesons have not been observed to date.
Baryonic decays of $B$ mesons provide good opportunities 
to study the mechanism of baryon production and to search for excited baryon
resonances~\cite{Mizuk:2004yu,Aubert:2006sp,Aubert:2007eb}.
The observation of intriguing behavior in the baryonic decays of the $\Bz$ and $\Bp$ mesons, \eg the enhancements of the rate of 
multi-body decays and the production of baryon pairs of low
mass~\cite{Lee:2004mg,Abe:2004sr,Aubert:2005gw,Medvedeva:2007af,Wei:2007fg,Chen:2008jy,delAmoSanchez:2011gi},
has further motivated this study.
\par
This Letter presents the first observation of a baryonic $\Bcp$ decay, $\DecayPsiPPH$, and
the measurement of its branching fraction with respect to the channel $\DecayPsiH$. 
The mass of the \Bcp meson is also determined using the $\DecayPsiPPH$ channel.
Owing to the small energy release ($Q$-value) of this channel, 
the systematic uncertainty of the measured \Bcp mass is small compared to the $\DecayPsiH$ channel. 
\par
The data used in this analysis are from $pp$ collisions recorded by the \lhcb experiment, 
corresponding to an integrated luminosity of $1.0\invfb$ at a center-of-mass energy of $7\tev$ 
and $2.0\invfb$ at $8\tev$.
The \lhcb detector~\cite{Alves:2008zz} is a single-arm forward
spectrometer covering the pseudorapidity range \mbox{$2<\eta <5$},
designed for the study of particles containing \bquark or \cquark
quarks. The detector includes a high-precision tracking system
consisting of a silicon-strip vertex detector surrounding the $pp$
interaction region, a large-area silicon-strip detector located
upstream of a dipole magnet with a bending power of about
$4{\rm\,Tm}$, and three stations of silicon-strip detectors and straw
drift tubes placed downstream~\cite{LHCb-DP-2013-003}.
The combined tracking system provides a momentum measurement with
relative uncertainty varying from 0.4\% at low momentum to 0.6\% at 100\gevc,
and impact parameter resolution of 20\mum for
tracks with large transverse momentum ($\pt$). Different types of charged hadrons are distinguished using information
from two ring-imaging Cherenkov detectors~\cite{LHCb-DP-2012-003}. 
Photon, electron and hadron candidates are identified by a calorimeter system consisting of 
scintillating-pad and preshower detectors, an electromagnetic calorimeter and a hadronic calorimeter.
Muons are identified by a
system composed of alternating layers of iron and multiwire
proportional chambers~\cite{LHCb-DP-2012-002}. The trigger~\cite{LHCb-DP-2012-004} consists of a hardware stage, based on
information from the calorimeter and muon systems, followed by a software stage, which
applies a full event reconstruction. 
In this analysis, \jpsi candidates are reconstructed in the dimuon
decay channel, and only trigger information related to the final state muons is considered.
Events are selected by the hardware triggers requiring a single muon 
with $\pt>1.48\gevc$ or a muon pair with product of transverse momenta 
greater than $(1.3\,\mathrm{GeV}\!/c)^2$. 
At the first stage of the software trigger, events are selected that contain two muon
tracks with $\pt>0.5\gevc$ and invariant mass $M(\mumu)>2.7\gevcc$, or a single muon track with $\pt>1\gevc$ and $\chi^2$ of 
the impact parameter ($\chi^2_\mathrm{IP}$) greater than 16 with respect to any primary vertices. 
The quantity $\chi^2_\mathrm{IP}$ is the difference between the $\chi^2$ values of a given primary vertex reconstructed with 
and without the considered track. The second stage of the software trigger selects a muon pair with an invariant 
mass that is consistent with the known $\jpsi$ mass~\cite{PDG2012}, 
with the effective decay length significance of the reconstructed $\jpsi$ candidate, 
$S_\mathrm{L}$, greater than 3, where $S_\mathrm{L}$ is 
the distance between the $\jpsi$ vertex and the primary vertex divided by its uncertainty.  

\par
The offline analysis uses a preselection,
followed by a multivariate selection based on a boosted decision tree (BDT)~\cite{Breiman,AdaBoost}.
In the preselection, the invariant mass of the $\jpsi$ candidate is required to be in the interval
$[3020,3135]\mevcc$.
The $\jpsi$ candidates are selected by requiring the $\chi^2$ 
per degree of freedom, $\chi^2/\mathrm{ndf}$, of the vertex fit to be less than 20.
The muons are required to have $\chi^2_\mathrm{IP}>4$ with respect to any reconstructed $pp$ vertex, to suppress 
the $\jpsi$ candidates produced promptly in $pp$ collisions. 
The decay $\DecayPsiH$ ($\DecayPsiPPH$) is reconstructed by combining a \jpsi candidate with
one (three) charged track(s) under $\pip$ ($p$, $\overline{p}$ and $\pip$) mass hypothesis. 
The requirements $\chi^2_\mathrm{IP}>4$ and $\pt>0.1\gevc$, are applied to these hadron tracks.
Particle identification (PID) is performed using dedicated neural networks, which use the information from all the
sub-detectors.
Well-identified pions are selected by a tight requirement on the value of the PID discriminant $\mathcal{P}_\pi$.
A loose requirement is applied to the PID discriminants of protons and
anti-protons, $\mathcal{P}_{p}$, $\mathcal{P}_{\overline{p}}$,  followed by the optimization described below.
To improve the PID performance, the momenta of protons and anti-protons are required to be
greater than $10\gevc$. The $\Bcp$ candidate is required to have vertex fit
$\chi^2/\mathrm{ndf}<6$, $\pt>2\gevc$, $\chi^2_\mathrm{IP}<16$ with respect to at least one 
reconstructed $pp$ collision and decay-time significance larger than 9 with respect to the vertex with the smallest
$\chi^2_\mathrm{IP}$.
To improve the mass and decay-time resolutions, a kinematic fit~\cite{Hulsbergen:2005pu} is applied to the $\Bcp$
decay, constraining the mass of the $\jpsi$ candidate to the current best world average~\cite{PDG2012} and the momentum of the $\Bcp$
candidate to point back to the primary vertex. 
\par
The BDT is trained with a simulated sample, where $\Bcp$
candidates are generated with \bcvegpy~\cite{Chang:2005hq}, interfaced
to $\textsc{Pythia6}$~\cite{Sjostrand:2006za}, using a specific \lhcb
configuration~\cite{LHCb-PROC-2010-056}. Decays of hadronic particles
are described by \evtgen~\cite{Lange:2001uf}, in which final-state
radiation (FSR) is generated using \photos~\cite{Golonka:2005pn}. The
interaction of the generated particles with the detector and its
response are implemented using the \geant
toolkit~\cite{Allison:2006ve, *Agostinelli:2002hh} as described in
Ref.~\cite{LHCb-PROC-2011-006}.  
For the background, candidates in the invariant mass sidebands
of the preselected $\Bcp$ data sample are used. The BDT input variables are $\pt$, $\chi^2_\mathrm{IP}$, $S_\mathrm{L}$ of the
$\Bcp$ candidate, $\chi^2/\mathrm{ndf}$ of its vertex fit, the quality of the constrained kinematic fit of the decay chain, and 
$\pt$, $\chi^2_\mathrm{IP}$ of the hadrons.
For the $\DecayPsiPPH$ candidates, the selection criteria are fixed by optimizing the BDT discriminant jointly
with the product of two proton PID discriminants, $\mathcal{P}_p\times\mathcal{P}_{\overline{p}}$. The selections on BDT
discriminant and the combined PID discriminant are chosen to maximize the figure of merit, aiming for a signal significance of three
standard deviations,
$\epsilon/(3/2+\sqrt{B})$~\cite{Punzi:2003bu}, where $\epsilon$ is the signal efficiency determined using simulated
events and  $B$ is the number of expected background
candidates estimated using sideband events in the data. 
For the $\DecayPsiH$ decay, the BDT discriminant is selected to maximize
the signal significance $S/\sqrt{S+B}$, where $S$ and $B$ are the expected signal and background yields,
estimated from simulated events and sideband data, respectively.
\par
\begin{figure}[!tbp]
  \begin{minipage}[t]{0.5\textwidth}
    \centering
    \includegraphics[width=1.0\textwidth]{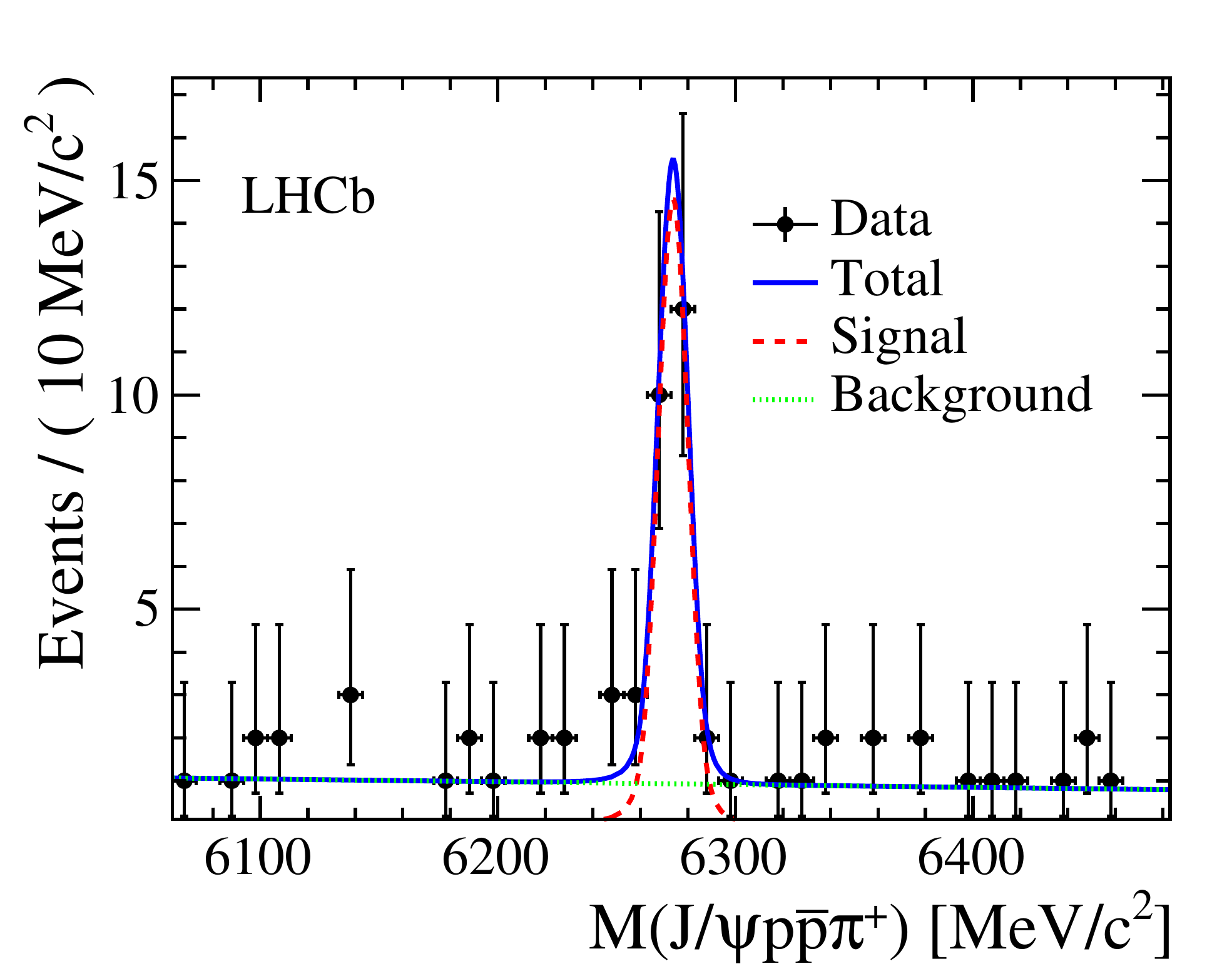}
  \end{minipage}
  \begin{minipage}[t]{0.5\textwidth}
    \centering
    \includegraphics[width=1.0\textwidth]{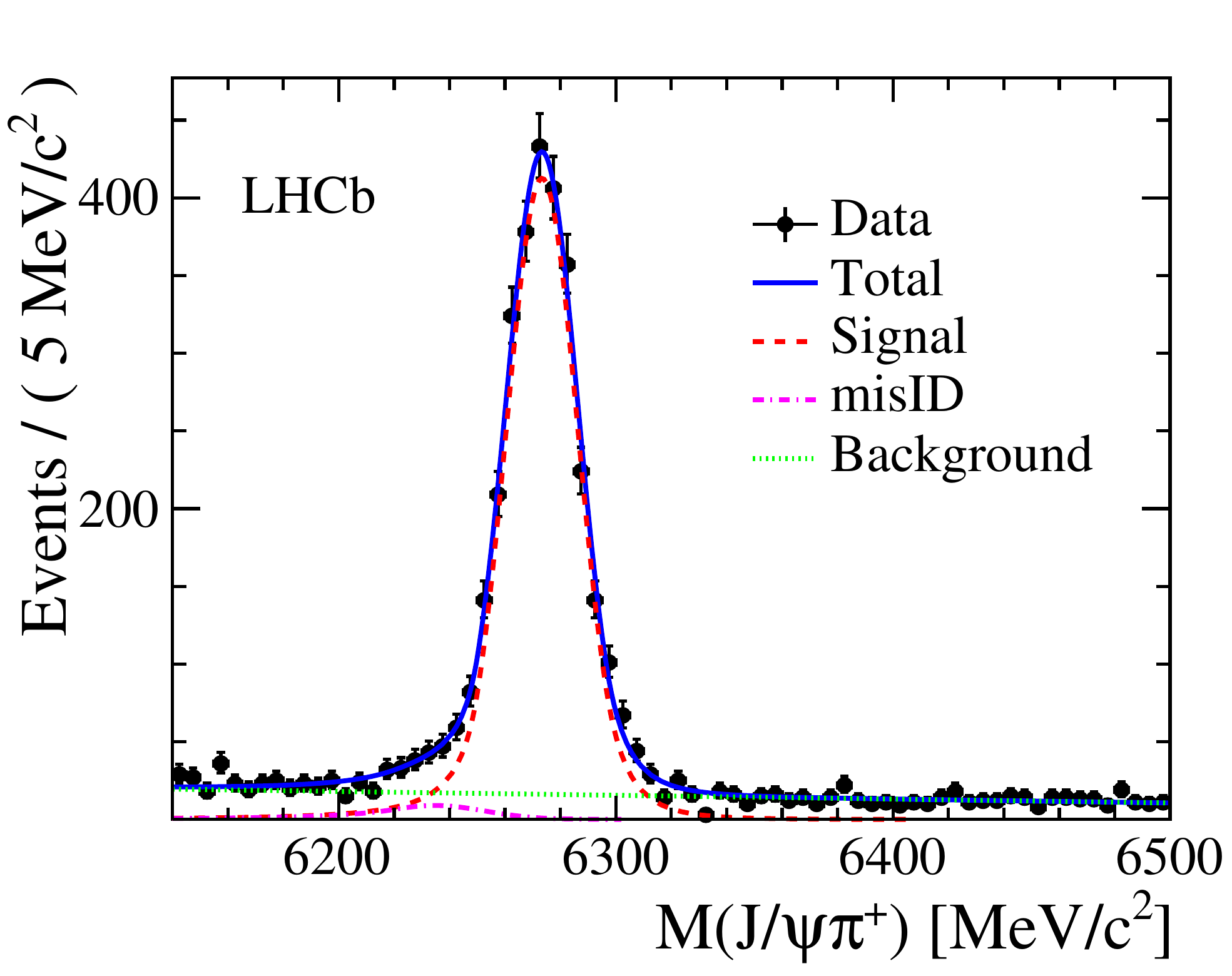}
  \end{minipage}
\caption{Invariant mass distribution for (left) $\DecayPsiPPH$ and (right)
$\DecayPsiH$ candidates.  The superimposed curves show the fitted contributions from signal (dashed), combinatorial background (dotted),
misidentification background (dot-dashed) and their sum (solid).}
  \label{fig:IM}
\end{figure}

Figure~\ref{fig:IM} shows the invariant mass distributions of the $\DecayPsiPPH$ and $\DecayPsiH$ candidates after all
selections, together with the results of unbinned extended maximum likelihood fits.
For both decays, the signal shape is modeled with a modified Gaussian distribution with power-law tails on both sides,
with the tail parameters fixed from simulation. The combinatorial background is described by a linear function.
The $\DecayPsiH$ channel is affected by a peaking background from the $\Bcp\to\jpsi\Kp$ decay where the kaon is misidentified as a
pion. The shape of this component is taken from the simulation and its yield, relative to the $\DecayPsiH$ decay,
is fixed to the ratio of their branching fractions, $0.069\pm0.019$~\cite{LHCb-PAPER-2013-021}, corrected by their relative efficiency.
The invariant mass resolution for the $\DecayPsiH$ decay is determined to be $13.0\pm0.3\mevcc$,
which is the width of the core of the modified Gaussian, and the value in the simulated sample is $11.69\pm0.06\mevcc$.
In the fit to the $\DecayPsiPPH$ invariant mass distribution, the signal resolution is fixed to
$6.40\mevcc$, which is the measured resolution of $\DecayPsiH$ decay in data scaled with their ratio in simulation,
$0.492\pm0.005\,(\mathrm{stat})$.
The observed signal yields are $23.9\pm5.3$ ($2835\pm58$) for the $\DecayPsiPPH$ ($\DecayPsiH$) decay, where the
uncertainties are statistical. 
The significance of the decay $\DecayPsiPPH$ is $7.3\,\sigma$, determined from the likelihood ratio of the fits with
background only and with signal plus background hypotheses~\cite{Cowan:2010js}. 

\par
From the fit to the \Bcp invariant mass distribution in the $\DecayPsiPPH$ decay, the mass of the \Bcp meson is found to
be $6273.8\pm1.8\mevcc$.
Table~\ref{tab:SysMM} summarizes the systematic uncertainties of the $\Bcp$ mass measurement,
which are dominated by the momentum scale calibration.
The alignment of the \lhcb tracking system is performed with samples of prompt $\Dz\to K^-\pip$ decays, 
and the momentum is calibrated using $\Kp$ from $\Bp\to\jpsi\Kp$ decays, and validated using a variety of known resonances. 
The uncertainty of the momentum scale calibration is $0.03\%$~\cite{LHCb-PAPER-2013-011}, which is the
difference between momentum scale factors determined using different resonances.
This effect is studied by changing the momentum scale by one standard deviation and repeating the analysis,
taking the variation of the reconstructed mass as a systematic uncertainty. The amount of
material traversed by a charged particle in the tracking system is known with an uncertainty of 10\%, and the systematic
effect of this uncertainty on the \Bcp mass measurement is studied by varying the energy loss correction by
10\% in the reconstruction~\cite{LHCb-PAPER-2011-035}. 
Since only charged tracks are reconstructed, the $\Bcp$ mass is underestimated due to FSR by $0.20\pm0.03\mevcc$,
as determined with a simulated sample. 
Therefore, the measured mass is corrected by $0.20\mevcc$ and $0.03\mevcc$ is assigned as a systematic uncertainty. The
contribution from the fit model is studied by using alternative fit functions for the signal and background, by using different fit invariant mass
ranges or by changing the estimated mass resolution within its uncertainty. 
The total systematic uncertainty of the mass measurement is $0.42\mevcc$.
After the correction for FSR, the mass of the \Bcp meson is determined to be
$6274.0\pm1.8\,(\mathrm{stat})\pm0.4\,(\mathrm{syst})\mevcc$. 
A combination of this result with previous \lhcb measurements
gives $6274.7\pm0.9\,(\mathrm{stat})\pm0.8\,(\mathrm{syst})\mevcc$. 
In the combination of the mass measurements, all systematic uncertainties apart from those due to the mass fit model and FSR 
are considered fully correlated.

\begin{table}[!tb]
\caption{Systematic uncertainties for the $\Bcp$ mass measurement.}
\label{tab:SysMM}
\centering
\begin{tabular}{lc}
\toprule[1.0pt]
Source&Value ($\mevcc$)\\
\midrule[1.0pt]
Momentum scale&0.40\\
Energy loss&0.05\\
Final state radiation&0.03\\
Fit model&0.10\\
\midrule[1.0pt]
Total&0.42\\
\bottomrule[1.0pt]
\end{tabular}
\end{table}
\par
In the branching fraction measurement of the decay $\DecayPsiPPH$, 
to account for any difference between data and simulation, 
the PID efficiency is calibrated using control data samples.
To allow easy calibration of the PID efficiency, the selection on the individual PID discriminants, 
$\mathcal{P}_p$ and $\mathcal{P}_{\overline{p}}$, is applied instead of their product. 
The same cut value is applied to the two PID variables,
and this cut value is optimized simultaneously with the BDT discriminant, maximizing the
same figure of merit.  With the new selection criteria, used to determine the branching fraction,
the signal yield of the $\DecayPsiPPH$ decay is $19.3^{\,+\,5.3}_{\,-\,4.6}\,(\mathrm{stat})$.
The ratio of yields between the $\DecayPsiPPH$ and $\DecayPsiH$ modes is determined to be
$r_N=0.0068^{\,+\,0.0019}_{\,-\,0.0016}\,(\mathrm{stat})$.
\par
The ratio of branching fractions is calculated as $$\frac{\BF(\DecayPsiPPH)}{\BF(\DecayPsiH)} = \frac{r_N}{r_\epsilon},$$
where $r_\epsilon\equiv\epsilon(\DecayPsiPPH)/\epsilon(\DecayPsiH)$ is the ratio of the total efficiencies. 
The geometrical acceptance, reconstruction, selection and trigger efficiencies are determined from simulated
samples for both channels. 
The central value of the \Bcp lifetime measured by \lhcb, $509\pm8\,(\mathrm{stat})\pm12\,(\mathrm{syst})\fs$~\cite{LHCb-PAPER-2013-063}, is used in the simulation. 
The PID efficiency for each track is measured in data in bins of 
momentum, $p$, pseudorapidity, $\eta$ of the track and track multiplicity of the event, $n_\mathrm{trk}$. 
The PID efficiency for pions is determined with $\pip$ from $D^{*}$-tagged $D^0\to \Km\pip$ decays. 
Similarly, the PID efficiency for protons is determined using protons from $\Lambda_c^+\to p\Km\pip$ decays. 
These efficiencies are assigned to the simulated candidate according to $p$ and $\eta$ 
of the final state hadron tracks, and $n_\mathrm{trk}$ of the event.
The distribution of $n_\mathrm{trk}$ in simulation is reweighted to match that in data.  
The overall ratio of efficiencies, $r_\epsilon$, is found to be $(4.76\pm0.06)\%$, where the uncertainty is statistical.  

\begin{table}[!tb]
\caption{Systematic uncertainties (in percent) for the relative branching fraction measurement.}
\label{tab:SysBr}
\centering
\begin{tabular}{lc}
\toprule[1.0pt]
Source&Value (\%)\\
\midrule[1.0pt]
Fit model&2.0\\
Acceptance&0.7\\
Trigger&1.1\\
Lifetime&1.1\\
Reco. of $p,\,\overline{p}$&$2\times2.3$\\
Pion PID&$1.1$\\
Proton PID&$2.4$\\
Decay model&$6.7$\\
\midrule[1.0pt]
Total&8.9\\
\bottomrule[1.0pt]
\end{tabular}
\end{table}
\par
\begin{figure}[!tb]
    \centering
  \begin{minipage}[t]{0.8\textwidth}
    \centering
    \includegraphics[width=1.0\textwidth]{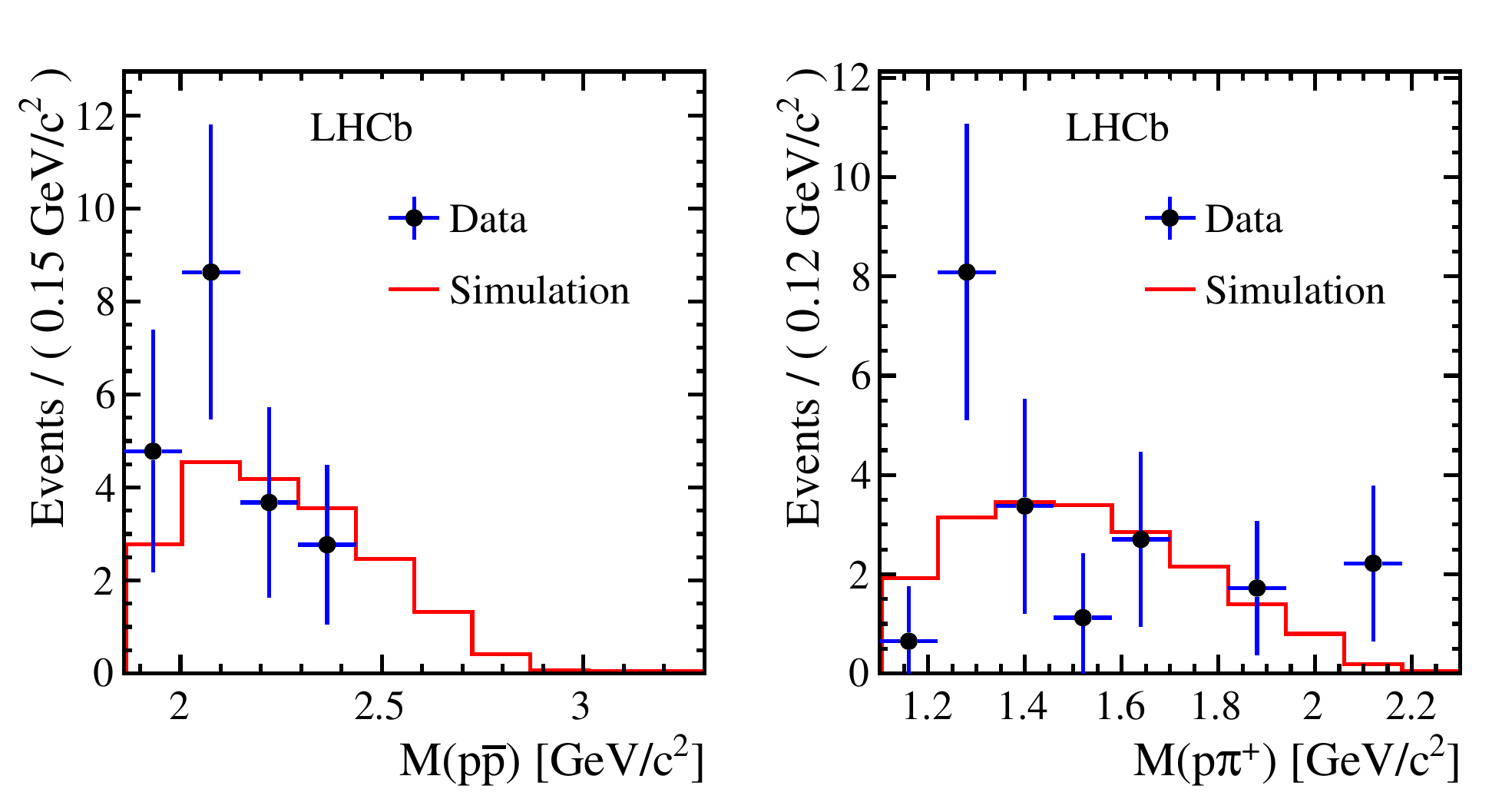}
  \end{minipage}
  \caption{Invariant mass distributions of (left) $M(p\overline{p})$ and (right) $M(p\pip)$ for
  data (dots) and simulation (solid) using uniform phase-space model, for \DecayPsiPPH decay. } 
  \label{fig:DecayModel}
\end{figure}

The systematic uncertainties for the branching fraction measurement are summarized in Table~\ref{tab:SysBr}. 
For the signal yields, the systematic uncertainty is obtained by varying the invariant mass fit functions of the two modes. 
The effect of geometrical acceptance is evaluated by comparing the efficiencies obtained from samples simulated with 
different data taking conditions.
The systematic uncertainty due to the trigger requirement is studied by comparing the trigger efficiency in data and simulated
samples, using a large $\jpsi$ sample~\cite{LHCb-PAPER-2012-010,LHCb-PAPER-2013-010}. 
The impact of the uncertainty of the $\Bcp$ lifetime is evaluated from the variation of the relative efficiency when the
$\Bcp$ lifetime is changed by one standard deviation of the \lhcb measurement~\cite{LHCb-PAPER-2013-063}. 
The systematic uncertainty associated with the reconstruction efficiency of the two additional hadron tracks, $p$ and $\overline{p}$, in the $\DecayPsiPPH$
mode compared to the $\DecayPsiH$ mode, is also studied.
Different assumptions for the pion PID efficiency in the kinematic regions where no calibration efficiency is available introduce a systematic uncertainty. 
For the protons, the systematic uncertainty from PID selection takes into account the uncertainties in the single-track efficiencies, the binning scheme
in $(p,\,\eta,\,n_\mathrm{trk})$ intervals and the uncertainty of the track multiplicity distribution. 
Another systematic uncertainty is related to the unknown decay model of the mode $\DecayPsiPPH$. 
The simulated sample is generated according to a uniform phase-space decay model. 
Figure~\ref{fig:DecayModel} shows the one-dimensional invariant mass distributions of $M(p\overline{p})$ and
$M(p\pip)$ for data, with background subtracted using the \sPlot method~\cite{Pivk:2004ty}. Figure~\ref{fig:DecayModel}
also shows the distributions for simulated events, which agree with those in data within the large statistical
uncertainties. 
The efficiency calculated using the observed distribution in data 
relative to the efficiency determined using the simulated decay model is $0.949\pm0.067$, where the uncertainty is statistical. 
Since the value is consistent with unity within the uncertainty, no correction to the efficiency is made and a
systematic uncertainty of $6.7\%$ is assigned. 
The total systematic uncertainty associated with the relative branching fraction measurement is $8.9\%$.  
\par
As a result the ratio of branching fractions is measured to be 
\begin{equation*}
\frac{\BF(\DecayPsiPPH)}{\BF(\DecayPsiH)} = 0.143^{\,+\,0.039}_{\,-\,0.034}\,(\mathrm{stat})\pm0.013\,(\mathrm{syst}),
\end{equation*}
which is consistent with the expectation from the spectator decay model assuming factorization~\cite{Cheng:2012fq}, 
$\frac{\BF(\DecayPsiPPH)}{\BF(\DecayPsiH)}\sim\frac{\BF(\decay{\Bd}{D^{*-} p \overline{p}\pip})}{\BF(\decay{\Bd}{D^{*-}\pip})}  = 0.17\pm0.02$.
The branching fractions for $\decay{\Bd}{D^{*-} p \overline{p}\pip}$ and $\decay{\Bd}{D^{*-}\pip}$ decays are taken from
Ref.~\cite{PDG2012}.
\par
In conclusion, the decay \DecayPsiPPH is observed with a significance of $7.3$ standard deviations, 
using a data sample corresponding to an integrated luminosity of $3.0\invfb$ collected by the \lhcb experiment.
Thi is the first observation of a baryonic decay of the \Bcp meson. The branching fraction of this decay 
relative to that of the $\DecayPsiH$ decay is measured.
The mass of the $\Bcp$ meson is measured to be $6274.0\pm1.8\,(\mathrm{stat})\pm0.4\,(\mathrm{syst})\mevcc$. 
In combination with previous results by \lhcb~\cite{LHCb-PAPER-2013-010, LHCb-PAPER-2012-028}, the
$\Bcp$ mass is determined to be $6274.7\pm0.9\,(\mathrm{stat})\pm0.8\,(\mathrm{syst})\mevcc$. 
\clearpage

\section*{Acknowledgements}
\noindent We express our gratitude to our colleagues in the CERN
accelerator departments for the excellent performance of the LHC. We
thank the technical and administrative staff at the LHCb
institutes. We acknowledge support from CERN and from the national
agencies: CAPES, CNPq, FAPERJ and FINEP (Brazil); NSFC (China);
CNRS/IN2P3 (France); BMBF, DFG, HGF and MPG (Germany); SFI (Ireland); INFN (Italy); 
FOM and NWO (The Netherlands); MNiSW and NCN (Poland); MEN/IFA (Romania); 
MinES and FANO (Russia); MinECo (Spain); SNSF and SER (Switzerland); 
NASU (Ukraine); STFC (United Kingdom); NSF (USA).
The Tier1 computing centres are supported by IN2P3 (France), KIT and BMBF 
(Germany), INFN (Italy), NWO and SURF (The Netherlands), PIC (Spain), GridPP 
(United Kingdom).
We are indebted to the communities behind the multiple open 
source software packages on which we depend. We are also thankful for the 
computing resources and the access to software R\&D tools provided by Yandex LLC (Russia).
Individual groups or members have received support from 
EPLANET, Marie Sk\l{}odowska-Curie Actions and ERC (European Union), 
Conseil g\'{e}n\'{e}ral de Haute-Savoie, Labex ENIGMASS and OCEVU, 
R\'{e}gion Auvergne (France), RFBR (Russia), XuntaGal and GENCAT (Spain), Royal Society and Royal
Commission for the Exhibition of 1851 (United Kingdom).

\addcontentsline{toc}{section}{References}
\setboolean{inbibliography}{true}
\bibliographystyle{LHCb}
\bibliography{main,LHCb-PAPER,LHCb-CONF,LHCb-DP,LHCb-TDR}

\newpage
\centerline{\large\bf LHCb collaboration}
\begin{flushleft}
\small
R.~Aaij$^{41}$, 
B.~Adeva$^{37}$, 
M.~Adinolfi$^{46}$, 
A.~Affolder$^{52}$, 
Z.~Ajaltouni$^{5}$, 
S.~Akar$^{6}$, 
J.~Albrecht$^{9}$, 
F.~Alessio$^{38}$, 
M.~Alexander$^{51}$, 
S.~Ali$^{41}$
% ...

{\footnotesize \it
% ...
$ ^{5}$Clermont Universit\'{e}, Universit\'{e} Blaise Pascal, CNRS/IN2P3, LPC, Clermont-Ferrand, France\\
$ ^{6}$CPPM, Aix-Marseille Universit\'{e}, CNRS/IN2P3, Marseille, France\\
% ...
$ ^{9}$Fakult\"{a}t Physik, Technische Universit\"{a}t Dortmund, Dortmund, Germany\\
% ...
$ ^{37}$Universidad de Santiago de Compostela, Santiago de Compostela, Spain\\
$ ^{38}$European Organization for Nuclear Research (CERN), Geneva, Switzerland\\
% ...
$ ^{41}$Nikhef National Institute for Subatomic Physics, Amsterdam, The Netherlands\\
% ...
$ ^{46}$H.H. Wills Physics Laboratory, University of Bristol, Bristol, United Kingdom\\
% ...
$ ^{51}$School of Physics and Astronomy, University of Glasgow, Glasgow, United Kingdom\\
$ ^{52}$Oliver Lodge Laboratory, University of Liverpool, Liverpool, United Kingdom\\
% ...
}
\end{flushleft}

\end{document}